\documentclass[aps,amsmath,amssymb,twocolumn,floatfix]{revtex4}

\usepackage{graphicx}
\usepackage{dcolumn}
\usepackage{bm}
\usepackage{color}
\usepackage{epsfig}
\usepackage{multirow}

\begin{document}

\preprint{Preprint}

\title{Electrostatic interaction in the presence of dielectric interfaces \\ and polarization-induced like-charge attraction}

\author{Zhenli Xu}\email{xuzl@sjtu.edu.cn}
\affiliation{Department of Mathematics, Institute of Natural Sciences, and MoE Key Lab for Scientific and Engineering Computing, Shanghai Jiao Tong University, Shanghai 200240, China}

\date{\today}

\begin{abstract}
Electrostatic polarization is important in many nano-/micro-scale physical systems such as colloidal suspensions,
biopolymers, and nanomaterials assembly. The calculation of polarization potential requires an efficient algorithm for solving 3D Poisson's equation. We have developed a useful image charge method to rapid evaluation of the Green's function of the Poisson's equation in the presence of spherical dielectric discontinuities. This paper presents an extensive study of this method by giving an convergence analysis and developing a coarse-graining algorithm. The use of the coarse graining could reduce the number of image charges to around a dozen, by 1-2 orders of magnitude.  We use the algorithm to investigate the interaction force between likely charged spheres in different dielectric environments. We find the size and charge asymmetry leads to an attraction between like charges, in agreement with existing results. Furthermore, we study three-body interaction and find in the presence of an external interface, the interaction force depends on the curvature of the interface, and behavior a non-monotonic electrostatic force.

\end{abstract}

\pacs{02.60.-x, 82.70.Dd, 77.22.Ej}
\keywords{ Green's function, Image charge method, Dielectric spheres, Like charge attraction}

\maketitle


\section{Introduction}

Electrostatic interaction is of fundamental importance in many fields of sciences at nano/micro scales. The long-range and many-body characters of the interaction lead to various challenges in the understanding of complex electrostatic phenomena \cite{Levin:RPP:02,FPPR:RMP:10,WKDG:NS:11}.  One debate in electrostatics is the like-charge attraction (LCA)  in colloidal science, which is not in agreement with the prediction based on the classical Poisson-Boltzmann (PB) theory. The LCA has been observed in a lot of experimental studies \cite{LG:Nat:97,GBPP:PT:00,TMV:SSC:08}, motivating wide computational and theoretical interest to understand Coulomb many-body phenomena (see \cite{GNS:RMP:02,FPPR:RMP:10,BKNNSS:PP:05} for reviews).

One of the central issues is the polarization effect due to the presence of dielectric interfaces. The dielectric constant of nano/microparticle materials could range from about 2 (hydrocabon objects) to the infinity (metallic objects), while that of surrounding environments could range from $1\sim2$ (air) to 80 (water) \cite{WKDG:NS:11}. The pairwise interaction potential and the ion distribution surrounding nanoparticles significantly depend on the dielectric properties of the system.  The effect of dielectric discontinuities has been often studied in various systems such as colloidal suspensions, liquid droplets in clouds, self-assembly of nanoparticles, and biopolymers.

The polarization due to interfaces remains a technical difficulty for particle-based computer simulations, and its algorithm development has attracted up-to-date attention \cite{JSD:PRL:12,GX:PRE:11,DBL:JCP:11,BBKS:JCP:10}. Direct numerical solutions of the Poisson's equation with finite element methods \cite{LZHM:CCP:08} are intensive for Monte Carlo (MC) and molecular dynamics (MD) simulations (for a survey of electrostatic algorithm in simulations, see \cite{AH:APS:05}), and analytical solutions are only available for simple geometries such as one spherical interface or cylindrical interface where harmonics series expansions or methods of image charges \cite{Thomson:JMPA:1845,Neumann:TL:83,OL:RSB:03,XC:SIREV:11,XCC:CCP:11} can be employed.
In simulations of electrolyte systems at room temperature, for instance, image effects has been widely studied in order to understand many-body electrostatic phenomena observed in colloidal suspensions or electric double layers \cite{TVP:JCP:82,TVO:JCP:84,KM:CPL:84,WK:JCP:06,WK:JCP:08,DBL:JCP:11}, and it could enhance the charge inversion of the colloid-microion complex at certain conditions \cite{WM:JCP:09,WM:JPCB:10,GXX:JCP:12}. For the LCA, it has been shown the entropy effect \cite{LD:JCP:90,JOL:JCP:06,JOL:JPCM:09,JL:JPCB:04,OJL:PRL:06,CDL:JCP:12,GGD:JCP:11} plays a main mechanism and it is not completely clear what is the role of image charges. This motivates us to develop a faster algorithm for Green's function problems in the presence of multiple spheres, used for simulations of such systems. The main emphasis of this study is the algorithm issues for the image potential of multiple spheres, suitable for general systems of different applications.  The MC simulations of colloid-colloid interactions in salt solutions by incorporating both the image charge and the entropy contributions will be discussed in a separate publication.

The electrostatic calculation between multiple dielectric spheres usually needs to use bi-spherical harmonic expansion \cite{Phil:JCP:74,MZ:JCIS:98,Linse:JCP:08,RL:JCP:08}.
Spherical harmonics are computationally expensive and slowly convergent for ions approaching to the surfaces in MC and MD simulations. Other methods have also been developed \cite{AH:JPCM:02,BBKS:JCP:10,LH:JCTC:06}. But, mostly analytical-based studies limit to no more than two spheres.  Alternatively, we have extended the discrete image approximation \cite{CDJ:JCP:07} of the Neumann's image principle to multiple spheres by reflections of discrete image charges among different interfaces  \cite{Xu:IS:12}. In this paper, we will present an extensive study and aspects such as the coarse-graining strategy and the convergence analysis will be proposed. Furthermore, we investigate the interaction force between likely charged spheres under size and charge asymmetries, and in the presence of an external interface, where the surface charge distribution is modeled by discrete charges.  The phenomenon of LCA at the conducting limit was widely known, and this work provides an efficient tool for a quantitative study of many-sphere interactions at finite dielectrics.

\section{Method}

We are interested in a system made of a cluster of dielectric spheres and a large number of ions with space charge distribution $\rho(\mathbf{r})=\sum_j q_j\delta(\mathbf{r}-\mathbf{r}_j)$, where $\delta(\cdot)$ is the Dirac delta, $\mathbf{r}_j$ and $q_j$ is the location and charge of the $j$th ion.
The electric potential $\Phi$ of the system satisfies the Poisson's equation,
\begin{equation}
-\nabla\cdot\varepsilon(\mathbf{r})\nabla\Phi(\mathbf{r})=4\pi\rho(\mathbf{r}).\label{poisson}
\end{equation}
We suppose the permittivity inside
the dielectric spheres is $\varepsilon_\mathrm{i}$ which could be varied for different spheres, and in the bulk solvent it is $\varepsilon_\mathrm{o}$,
then $\varepsilon(\mathbf{r})$ is piecewise constant.
Eq. \eqref{poisson} implies the boundary conditions on the spherical surfaces, namely, the electric potential
$\Phi$ and the electric displacement $\varepsilon(\mathbf{r})\partial\Phi(\mathbf{r})/\partial \mathbf{n}$
are continuous across the dielectric interfaces.

Suppose the electric potential decays to zero at the bulk solvent far away from the dielectric spheres. The solution of Eq. \eqref{poisson}  can be written as  \cite{Jackson:book:01},
\begin{equation}
\Phi(\mathbf{r})=\int G(\mathbf{r},\mathbf{r}') \rho(\mathbf{r}')d\mathbf{r}'=\sum_j q_jG(\mathbf{r}, \mathbf{r}_j), \label{green}
\end{equation}
where $G(\cdot,\cdot)$ is called the Green's function of the Poisson's equation, described by,
\begin{equation}
-\nabla\cdot\varepsilon(\mathbf{r})\nabla G(\mathbf{r},\mathbf{r}')=4\pi\delta(\mathbf{r}-\mathbf{r}').
\end{equation}
Physically, the Green's function is the electric potential produced by a unit point charge at the source point $\mathbf{r}'$.
Once the Green's function is explicitly given, we obtain the solution of the Poisson's equation for
an arbitrary charge distribution by Eq. \eqref{green}. As source charges are located outside spheres or on the interfaces, we rewrite the Green's function outside the dielectric spheres as the sum of the Coulomb potential and the polarization potential,
\begin{equation} G(\mathbf{r},\mathbf{r}')=\Phi_\mathrm{coul}(\mathbf{r},\mathbf{r}')+\Phi_\mathrm{pol}(\mathbf{r},\mathbf{r}'),
\end{equation}
with $\Phi_\mathrm{coul}(\mathbf{r})=1/\varepsilon_\mathrm{o}|\mathbf{r}-\mathbf{r}'|,$ the Green's function
in free space.

\subsection{Method of image charges}

\begin{figure}[htpb!]
\centering\includegraphics[scale=.35]{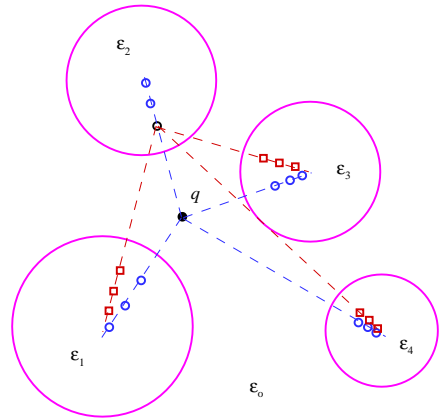}
\caption{Illustration of the construction of image charges by reflections among surfaces. A source charge $q$ produces the 1st-level
image charges inside each dielectric sphere, symbolled by blue circles and one black circle.
Each image produces the 2nd-level image charges inside other spheres, symbolled by red rectangles for
the black 1st-level image (Only this image's 2nd-level image charges are shown). The series is convergent by recursively performing this procedure.  } \label{schem}
\end{figure}

For the problem of a unique dielectric sphere in the system, it is known the Green's function solution can be obtained by the spherical harmonics series in the spherical
coordinates \cite{Jackson:book:01} or by the image charges, and our discussion focus on the later one. Suppose $R$ is the radius of the sphere,
$r$ and $r'$ are radial distances of $\mathbf{r}$ and $\mathbf{r}'$, and the spherical center
is the coordinate origin. The Neumann's image principle \cite{Neumann:TL:83} expresses the polarization potential as a line integral (See also \cite{CDJ:JCP:07,GX:PRE:11} for recent work),
\begin{equation}
\Phi_\mathrm{pol}=\frac{1}{\varepsilon_\mathrm{o}}
\int_0^{r_\mathrm{K}}\frac{f(r_\mathrm{K}/x)}{R |\mathbf{r}-\mathbf{x}|}dx,  \label{line}
\end{equation}
with
\begin{equation}
f(t)=-\gamma\delta(t-1)+\frac{\gamma t^\sigma}{\varepsilon+1}H(1-t).
\end{equation}
Here $\varepsilon=\varepsilon_\mathrm{i}/\varepsilon_\mathrm{o}$,
$\gamma=(\varepsilon_\mathrm{i}-\varepsilon_\mathrm{o})/(\varepsilon_\mathrm{i}+\varepsilon_\mathrm{o})$
and $\sigma=\varepsilon_\mathrm{o}/(\varepsilon_\mathrm{i}+\varepsilon_\mathrm{o})$
are three real constants, and $\mathbf{r}_\mathrm{K}=r_\mathrm{K}\mathbf{r}'/r'$,
$r_\mathrm{K}=R^2/r'$, and $\mathbf{x}=x\mathbf{r}'/r'$. It can be seen the Dirac delta function in the line charge density corresponds to
a point charge called the Kelvin image at $\mathbf{r}_\mathrm{K}$, and the Heaviside function $H(\cdot)$ corresponds to a continuous line charge.
The integral is from the origin to the point of the Kelvin image along the radial direction. A Gauss-Legendre quadrature to the
continuous line integral leads us to discrete image charges for approximating the polarization potential \cite{CDJ:JCP:07},
\begin{equation}
\Phi_\mathrm{pol}(\mathbf{r},\mathbf{r}')=\frac{1}{\varepsilon_\mathrm{o}}\sum_{m=1}^M\frac{q_m}{|\mathbf{r}-\mathbf{x}_m|}, \label{onesphere}
\end{equation}
where $q_1=-\gamma R/r'$ and $\mathbf{x}_1=\mathbf{r}_K$ are the parameters of the Kelvin image charge, and
$q_m=\omega_m\gamma R/(2r')$ and  $x_m=r_\mathrm{K}\left\{(1-s_m)/2\right\}^{1/\sigma},$
for $m=2,\cdots,M,$ are the charges and positions of the discrete point images due to the line integral contribution.
The coefficients $\{\omega_m, s_m, m=2,3,\cdots,M\}$ are the $(M-1)$-point Gauss weights and locations on the interval $[-1,1]$,
which can be found in many literature; for example, Numerical Recipes \cite{PTVF:book:92}. Besides the Gauss quadrature,
the integral can be approximated by a hypergeometric function \cite{DBL:JCP:11}, and historically other approximate
methods of images \cite{Friedman:MP:75,AT:JMB:94,Linse:JPC:86} are also developed.

When multiple dielectric interfaces  are present in the system, the polarization potential is constructed by
reflections between different surfaces, as is illustrated
in Figure \ref{schem}. Suppose we have $\mathcal{N}$
dielectric spheres. First of all, for a source charge, $M$ image point charges are
generated in each sphere by using the image expression for one sphere, Eq. \eqref{onesphere}. The potential
due to the summation of the source charge and the image charges in a specified sphere satisfies the interface
conditions on this spherical surface, but fails to satisfy the interface conditions on
the other spherical surfaces, and therefore $M$ new image charges in
each of the other sphere have to be used for each image charge. This procedure is recursively performed (see Figure \ref{schem}), leading to the following expression for the polarization potential of the whole system:
\begin{widetext}\begin{equation}
\Phi_\mathrm{pol}=\frac{1}{\varepsilon_\mathrm{o}}\sum_{j_1=1}^{\mathcal{N}}\sum_{m_1=1}^M
\left\{
\frac{q_{j_1m_1}}{|\mathbf{r}-\mathbf{x}_{j_1m_1}|}+\sum_{j_2=1 \atop j_2\neq j_1}^{\mathcal{N}}\sum_{m_2=1}^M
\left[
\frac{q_{j_2m_2j_1m_1}}{|\mathbf{r}-\mathbf{x}_{j_2m_2j_1m_1}|}+\sum_{j_3=1\atop j_3\neq j_2}^{\mathcal{N}}\sum_{m_3=1}^M
\left(
\frac{q_{j_3m_3j_2m_2j_1m_1}}{|\mathbf{r}-\mathbf{x}_{j_3m_3j_2m_2j_1m_1}|}+\cdots
\right)\right]\right\}, \label{allimage}
\end{equation}\end{widetext}
where the subscript ``$j_km_kj_{k-1}m_{k-1}\cdots j_1m_1$" refers to the next-level image charges of the charge with
subscript ``$j_{k-1}m_{k-1}\cdots j_1m_1$".

In Eq. \eqref{allimage}, there are $\mathcal{N}M$ image charges in the first term, which is double sums and we call
they are the images at the first level. The $n$th term is called the $n$th  level, which is $2n$-fold sums
and with $\mathcal{N}(\mathcal{N}-1)^{n-1}M^n$ images.  We truncate the series at $L$th level,
$
\Phi_\mathrm{pol}\approx\Phi_1+\Phi_2+\cdots+\Phi_L,
$
where $\Phi_i$ is the $2i$-fold sums for images at the $i$th level, then the total number of image charges is,
\begin{equation}
\frac{\mathcal{N}M^{L+1}(\mathcal{N}-1)^L-\mathcal{N}M}{\mathcal{N}M-M-1}. \label{imagenumber}
\end{equation}
A simple truncation of the series may lead to low
accuracy and expensive computation. It could be noticed that, in the one-sphere problem
the Kevin image charge has an opposite sign of the image line density, and
overall they are electric-neutral and behavior like a dipole. Higher levels give higher order multipoles, and presumedly
the series \eqref{allimage} has a very fast convergence.

All image charges are concentrated inside spheres with a high density. It is favorable to
combine some of them together by a coarse-graining strategy. This will greatly reduce the number of image charges,
as is discussed later on.

\subsection{Convergence analysis}

The dependence of the convergence of the image expression \eqref{allimage} on $M$ can be estimated by the truncation error of the Gauss quadrature which has a fast convergence. Usually two to three discrete points will provide a high accuracy to approximate the line integral \cite{CDJ:JCP:07,GX:PRE:11}. In this section, our analysis focuses on the dependence on $L$, the levels of reflections.

We consider a special case that two spheres of the same radius $R$ with separation $d=r_{12}-2R$ and $r_{12}$ being the distance between two centers.  The convergence rate as function of $\epsilon=d/R$ is estimated. We calculate the self energy of a unit source charge, which is equivalent to calculating the polarization potential at the source point.  The source charge is placed at the middle of two spheres, and the convergence is faster if the source charge deviates from this point. For asymmetric spheres, the following estimation still holds by taking $\epsilon=d/R_>$ where $R_>$ is the larger radius. Therefore we need only do the analysis for the special case to get a lower bound of the convergence rate.

We take $M=2$ so each reflection of a charge results in an image dipole. Suppose an image charge $q_I$ at $r_\mathrm{I}$
distance to the center of the sphere including itself. Let $r_s=r_{12}-r_\mathrm{I}$ and $\nu=(1/2)^{1/\sigma}$, then the potential contribution
of this image charge at the source site is,
\begin{equation}
\Phi_\mathrm{I}=\frac{q_I/\varepsilon_\mathrm{o}}{r_s-R-d/2}.
\end{equation}
Then this image's image dipole within the other sphere have the potential contribution,
\begin{equation}
\Phi_\mathrm{II}=\frac{-\gamma R q_\mathrm{I}}{r_s \varepsilon_\mathrm{o}}\cdot
\left(\frac{1}{R+\frac{d}{2}-\frac{R^2}{r_s}}-\frac{1}{R+\frac{d}{2}-\nu\frac{R^2}{r_s}}\right).
\end{equation}
Let us define $y=r_s/R$, then recalling $\epsilon=d/R$ we have,
\begin{equation}
\frac{\epsilon^2+4\epsilon+2}{2+\epsilon}\leq y\leq 2+\epsilon. \label{yrange}
\end{equation}
We calculate the absolute of the ratio between two potentials $\vartheta=\Phi_\mathrm{I}/\Phi_\mathrm{II}$, which is
\begin{equation}
|\vartheta|=\frac{\left[\left(1+\frac{\epsilon}{2}\right)y-1\right]\left[\left(1+\frac{\epsilon}{2}\right)y-\nu\right]}
{|\gamma|\left(1-\nu\right)\cdot\left(y-1-\frac{\epsilon}{2}\right)}.
\end{equation}
By using inequalities \eqref{yrange}, a simple derivation gives us an estimation of the convergence rate,
\begin{equation}
|\vartheta| \geq
\frac{2\eta\sqrt{(\eta^2-1)(\eta^2-\nu)}+\eta\left(2\eta^2-1-\nu\right)}{|\gamma|\left(1-\nu\right)} \label{rate}
\end{equation}
where $\eta=1+\epsilon/2.$

The charges at the $l$-th level are the images of those at the $(l-1)$-th level, thus Eq. \eqref{rate} provides a lower bound of the reflection at every level. We therefore obtain a convergence estimation of the truncation error, $\sim 1/|\vartheta|^{L}$, with respect to the reflection number $L$.

And at a certain separation distance, for example, $\epsilon\ge 1$,  the convergence of the image approximation is
fast. At the limit of infinite radius, $\epsilon=0$ and $\eta=1$ and the convergence rate is bigger than a constant value, $1/|\gamma|$.

\subsection{The coarse-graining algorithm}

After a couple of reflections, mostly image charges will be within the half radius of each sphere. We develop a coarse-graining technique to reduce the number of image charges, which is useful for dynamical simulations.

When the image system of a source has been generated from Eq. \eqref{allimage}, we do the following procedure by comparing all pairs of image charges inside every sphere.
Let $\delta>0$ be a given error criteria. Suppose $\mathbf{r}_{j,m}$ and $\mathbf{r}_{j,n}$ are the location of two image charges due to the source $j$ within sphere $k$, which has center $\mathbf{O}_k$ and radius $R_k$. If they both have radial distances less than half the radius, i.e., $|\mathbf{r}_{j,m}-\mathbf{O}_k|$ and $|\mathbf{r}_{j,n}-\mathbf{O}_k|<0.5R_k$, we calculate the distance between two points
\begin{equation}d_{mn}=|\mathbf{r}_{j,m}-\mathbf{r}_{j,n}|.\end{equation}
This distance is compared with the product of the distance criteria and the radius, and if
$d_{mn}<R_k\delta$, then we combine these two charges together, and place the new image charge at the weighted center of the two charges. Let $w_m=|q_{j,m}|/(|q_{j,m}|+|q_{j,n}|)$ and  $w_n=1-w_m$,
then the charge position is at $w_m\mathbf{r}_{j,m}+w_n\mathbf{r}_{j,n}$, and the charge strength is trivially the sum of two charges, $q_{j,m}+q_{j,n}$.

In the worst case, the coarse-graining algorithm has the error $\sim O(\delta)$ for the polarization potential which is independent of the radius of the sphere. However, this estimate is conservative if we recall the polarization potential is a dipole effect in global.
The performance of the algorithm will be discussed in the following section.

\section{Results}

In this section, the interaction between charged spheres with different dielectric permittivities is investigated.
We suppose the surrounding environment of the spheres is a homogeneous dielectric, which can be considered as a mimic of highly dilute solutions or a dielectric material in vacuum.

We consider two spheres of radii $R_1$ and $R_2$ with scaled separation $d=(r_{12}-R_1-R_2)/R_1$, each has $N=60$ charges on its surface.
The electrostatic energy of the system is composed of source-source interactions and the source-image interactions,
\begin{equation}
U(d)=\frac{1}{2\varepsilon_\mathrm{o}} {\sum_{i,j=1}^N}' \left( \frac{q_iq_j}{r_{ij}} + \sum_k \frac{q_i q_{j,k}}{|\mathbf{r}_i-\mathbf{r}_{j,k}|} \right)
\end{equation}
where $k$ ranges over all images of the $j$th source charge, and the image strengths and locations depend on the separation distance. The prime at the northeast of the summation represents that when $i=j$ the source-source interaction and the interaction between source and its first few images which do not change with $d$ should be not included. We define the interaction force between spheres by the negative gradient of the energy along the separation,
\begin{equation}
F(d)=-\partial U(d)/\partial d.
\end{equation}
Positive value of $F$ represents a repulsion between two particles, and negative one represents an attraction.
We use dimensionless unit by setting $\varepsilon_\mathrm{o}=1$,  the radius $R_1=1$ and interfacial ion valence $q_i=Q_1=1/N$ for the first sphere. The system asymmetries are described by $\varepsilon=\varepsilon_\mathrm{i}/\varepsilon_\mathrm{o}$, $R=R_2/R_1$ and $Q=Q_2/Q_1.$

\begin{figure}[htbp!]
\centering\includegraphics[scale=.25]{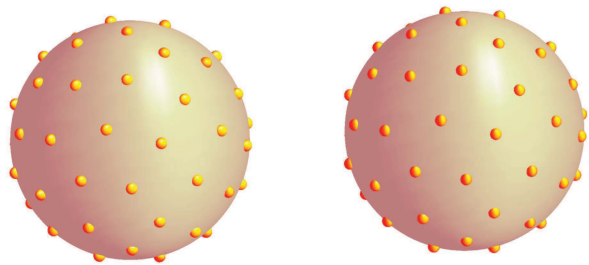}
\caption{The ion distribution on the surface of two spheres, generated by minimizing the electrostatic energy. } \label{twospheres}
\end{figure}

The configuration of interfacial ions are generated from an MC simulation of Coulomb repulsive particles on the surface. With the zero-temperature limit, the particles form a quasi Wigner-crystal structure, which is considered the ions are uniformly embedded on the surface in our study.  The configuration of the second sphere is a rotation of the first one, which minimizes the interaction energy of the ions between two spheres, and the ion distribution does not change with the varying of the interspherical separation. Figure \ref{twospheres} depicts the ion distribution on the two surfaces.

\subsection{Validation of accuracy for different parameters}

We let $R=1$, $Q=1$ and two sets of dielectric ratios $\varepsilon=0.02$ and 20.  These sets of systems are used to test the convergence and accuracy of the approximation with image charges. We compare the solutions between different numbers of discrete points for an image line, $M$, and deeps of reflection levels, $L$.

\begin{figure}[htbp!]
\centering\includegraphics[scale=.32]{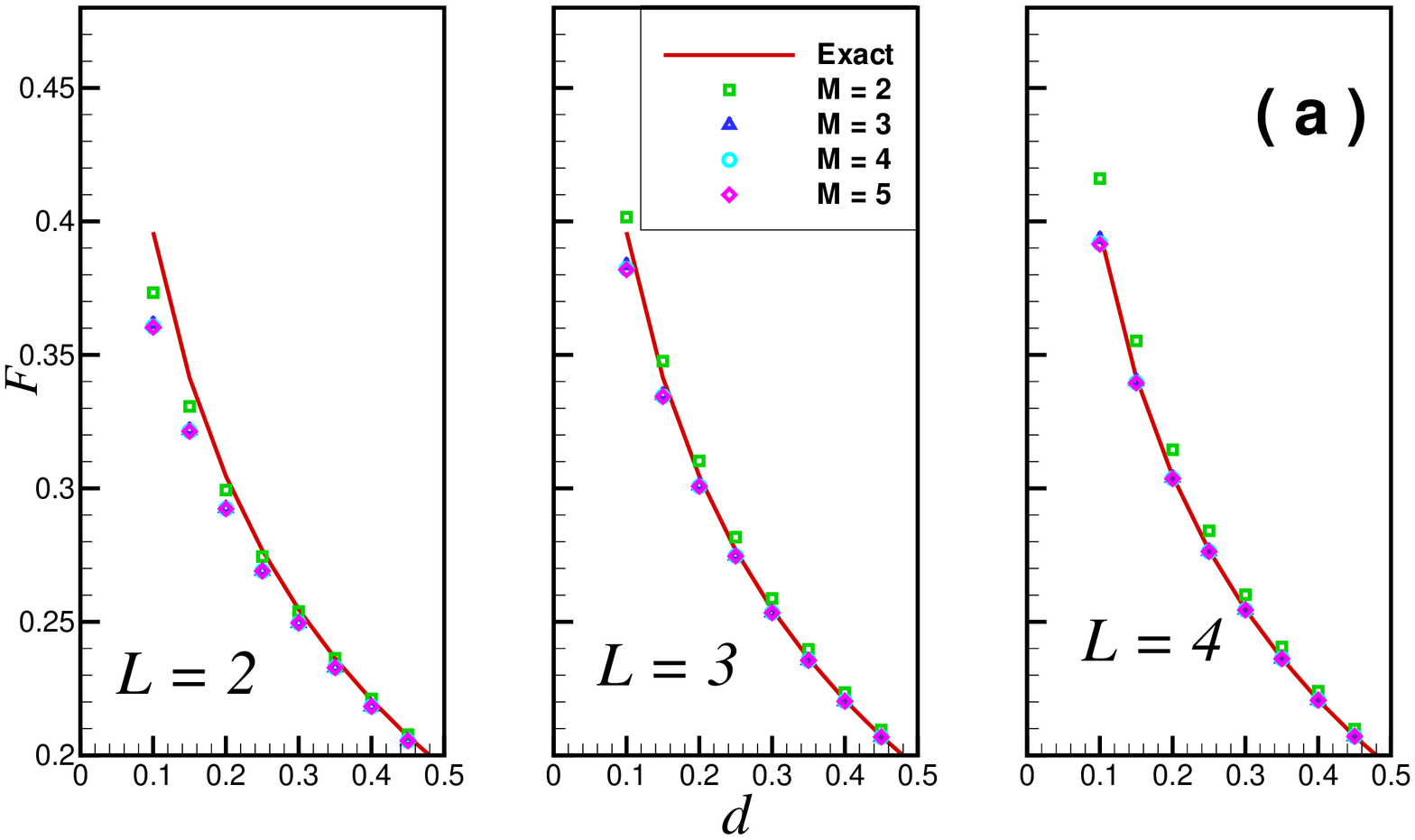} \\ \includegraphics[scale=.32]{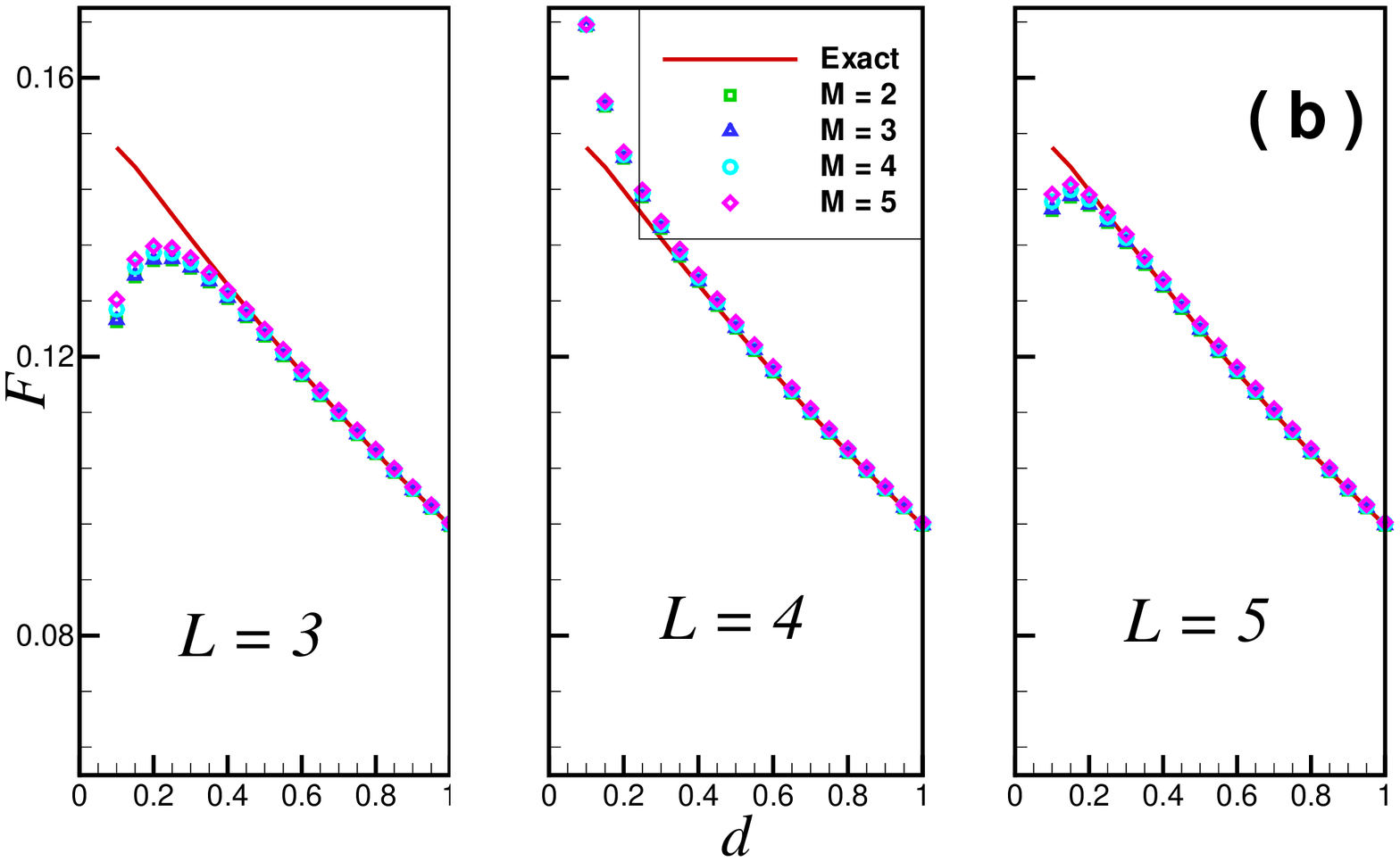}
\caption{Convergence of the interaction force between two charged spheres as a function of separation. Solutions with large $L$ are used for the ``exact" solutions. $R=1,$ and $Q=1.$ (a) $\varepsilon=0.02$; (b) $\varepsilon=20.$} \label{LCAconv}
\end{figure}
\begin{figure}[htbp!]
\centering\includegraphics[scale=.45]{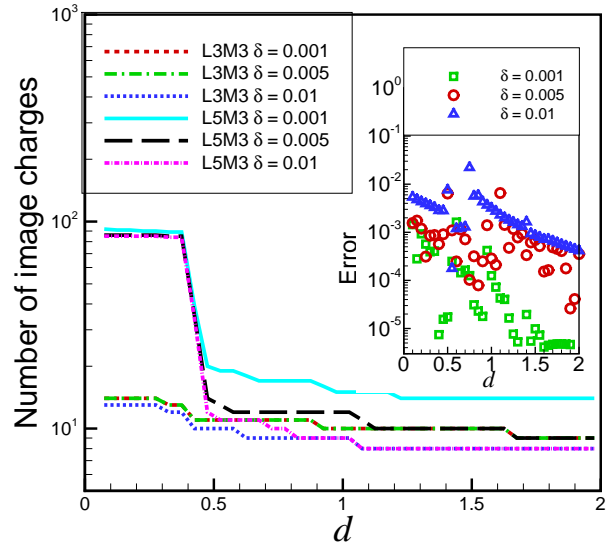}
\caption{Mean number of image charges of each source with the coarse-graining algorithm. The image numbers for $L3M3$ and $L5M3$ without the coarse graining are 78 and 726, respectively. The inset plot shows the relative errors of the $L5M3$ forces with the coarse graining in comparison to those of original image charges.} \label{CGfig}
\end{figure}

The results are illustrated in Figure \ref{LCAconv}, where the ``exact" is the reference solution from large $L$ and $M.$  It is seen all curves are very accurate for a separation $d>0.5$. For small $d$, however, more images should be used. The approximation has a fast convergence to $M$, and $M=2$ is already a useful accuracy since the charge neutrality is satisfied. In regard to the parameter $L$, the convergence has a dependence on $\varepsilon$. For small $\varepsilon$, the image solution converges rapidly with $L$. For large $\varepsilon$, an odd $L$ gives a fast convergence. This makes sense because the Kelvin image shares the same sign as the source for $\varepsilon<1$. While for $\varepsilon>1,$ the Kelvin image has an opposite sign, requiring the truncation at odd $L$.

To test the performance of the coarse-graining algorithm, we calculate the same system as in Figure \ref{LCAconv} for the case $\varepsilon=20$, and two sets of parameters $L=3, M=3$ and $L=5, M=3$, denoted by $L3M3$ and $L5M3$, which lead to 78 and 726 image numbers, respectively. We take $\delta=0.001, 0.005$ and $0.01$. The number of image charges and the relative errors of the force deviated from the original image charges as a function of the separation $d$ is illustrated in Figure \ref{CGfig}.  The coarse graining reduces the numbers of $L3M3$ and $L5M3$ to around 15 and 90, decreased by a factor of 5 and 8, respectively, in the case of $d<0.4$. The image number is further decreased for a larger $d$, as is shown that most of curves drop down to less than 10 with the increase of $d$.
For the relative errors of the forces of $L5M3$ with the coarse graining compared to those of original image charges, all three $\delta$ gives high accuracy, generally, less than 1\%, except a point at $d=0.725$ for $\delta=0.01$.

In the following calculations, $L5M3$ will be used, for which the image method has a reasonable accuracy for both small and large dielectric contrast.

\subsection{Size and charge asymmetries}

Many literature has documented that likely charged spheres attract one each other due to the polarization under different size and charge, and different dielectric contrast. In a recent work, Lekner \cite{Lekner:PRSA:12} demonstrated that at small separation two conducting spheres almost always attract each other. For finite dielectrics, Bichoutskaia {\it et al.} \cite{BBKS:JCP:10} calculated the dependence of interaction force on size, charge, and dielectric ratios; see also references therein for an overview of historic works. Most of existing calculations use uniform representation of the surface charge, instead of the discrete representation in our work.

\begin{figure*}[htbp!]
\centering\includegraphics[scale=.32]{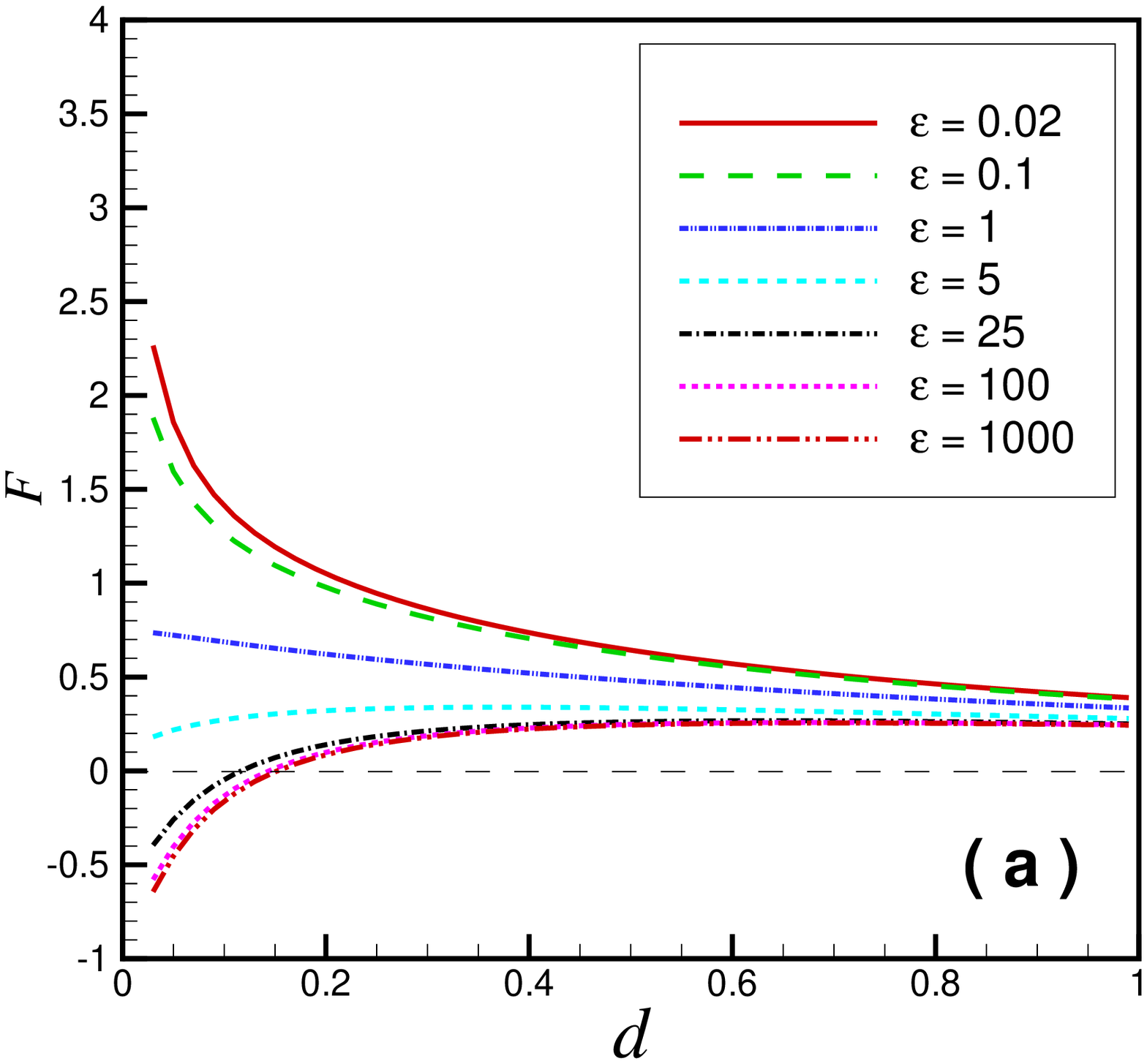} \includegraphics[scale=.32]{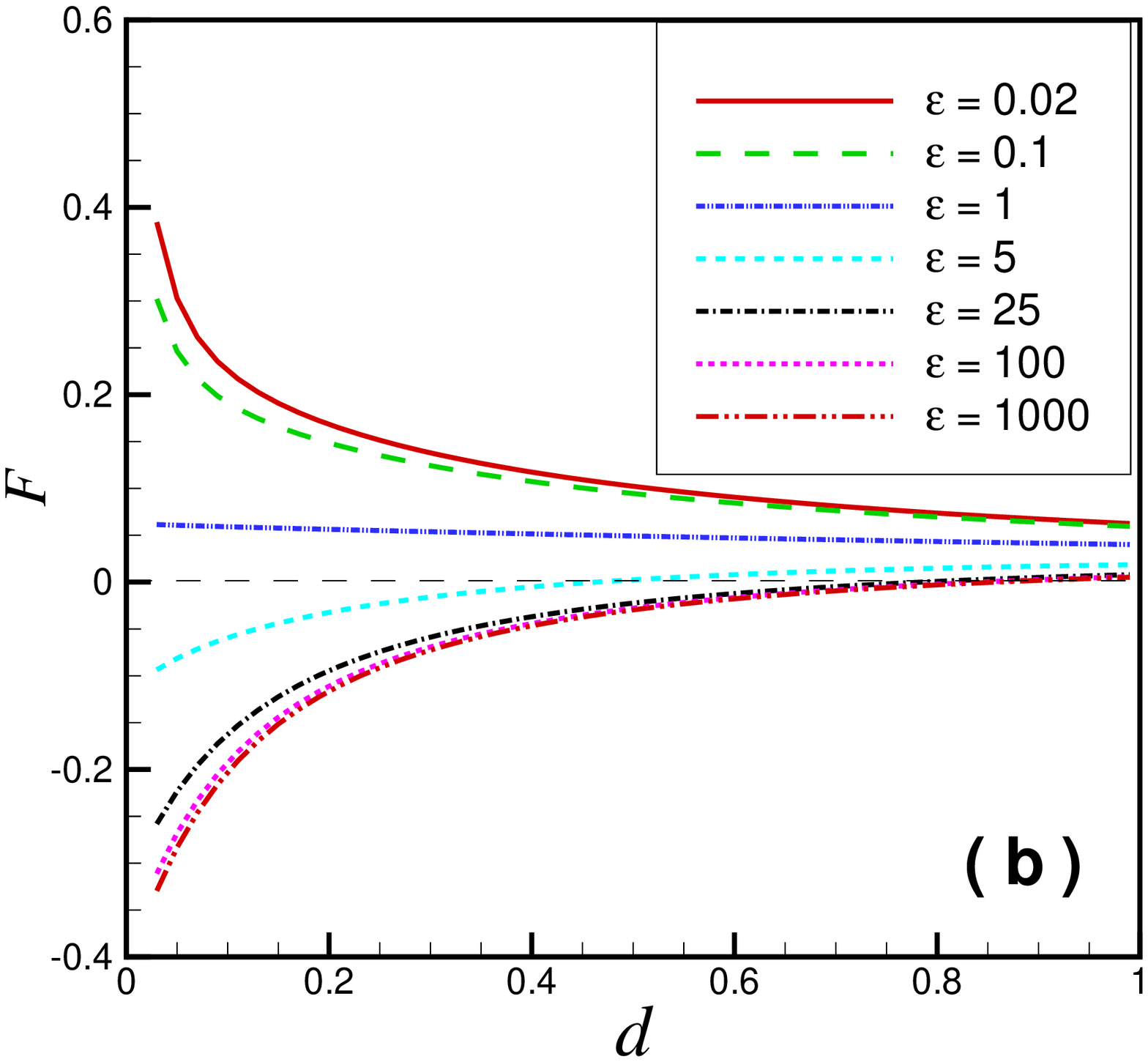}
\caption{Interaction force for asymmetric like charges as a function of separation $d$. The spheres are attractive at small separation and large $\varepsilon$.  The parameters are: (a) $R=1$, $Q=3;$  (b)  $R=3$, $Q=1.$  } \label{ChaSizeAsy}
\end{figure*}

We calculate the force for varying dielectric ratios by taking two groups of settings: one is with $R=1$ and $Q=3$, and the other one with $R=3$ and $Q=1$.  The results with size and charge asymmetries for dielectric contrast $\varepsilon$ varying from 0.02 to 1000 are shown in Figure \ref{ChaSizeAsy}. The plots demonstrate the breaking of symmetry does lead to an attraction between likely charged spheres at small separation. When $\epsilon>25$, the dielectric spheres have very similar property as conductor. In the right panel of the results, the attraction regime for high dielectric contrast is about $R$, which is because the second sphere has a radius of $3R$ and thus the image dipole in this sphere has longer-ranged interaction.

\subsection{Three-body interaction}

\begin{figure*}[htbp!]
\vspace{0.3cm}
\centering\includegraphics[scale=.3]{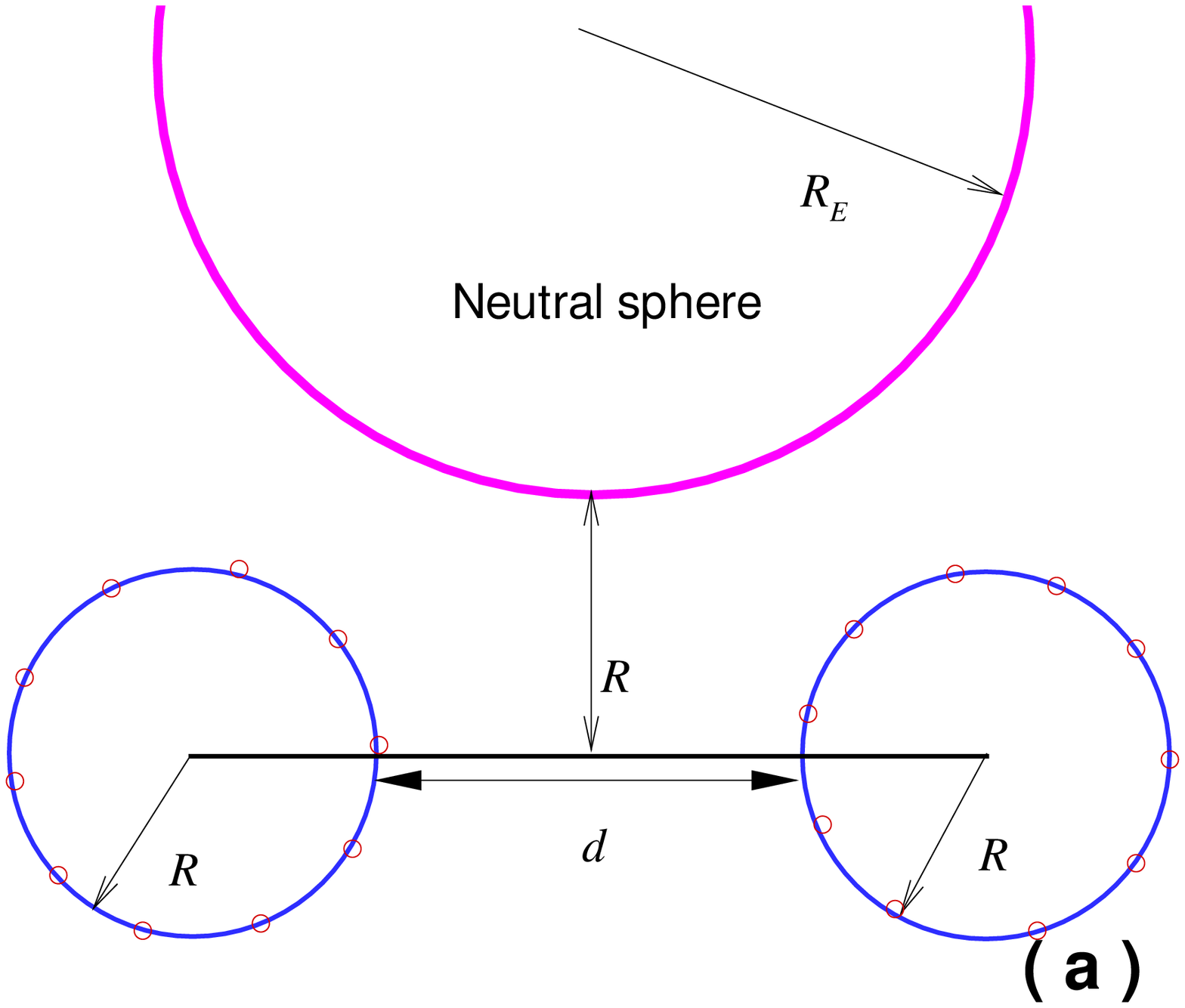}\includegraphics[scale=.3]{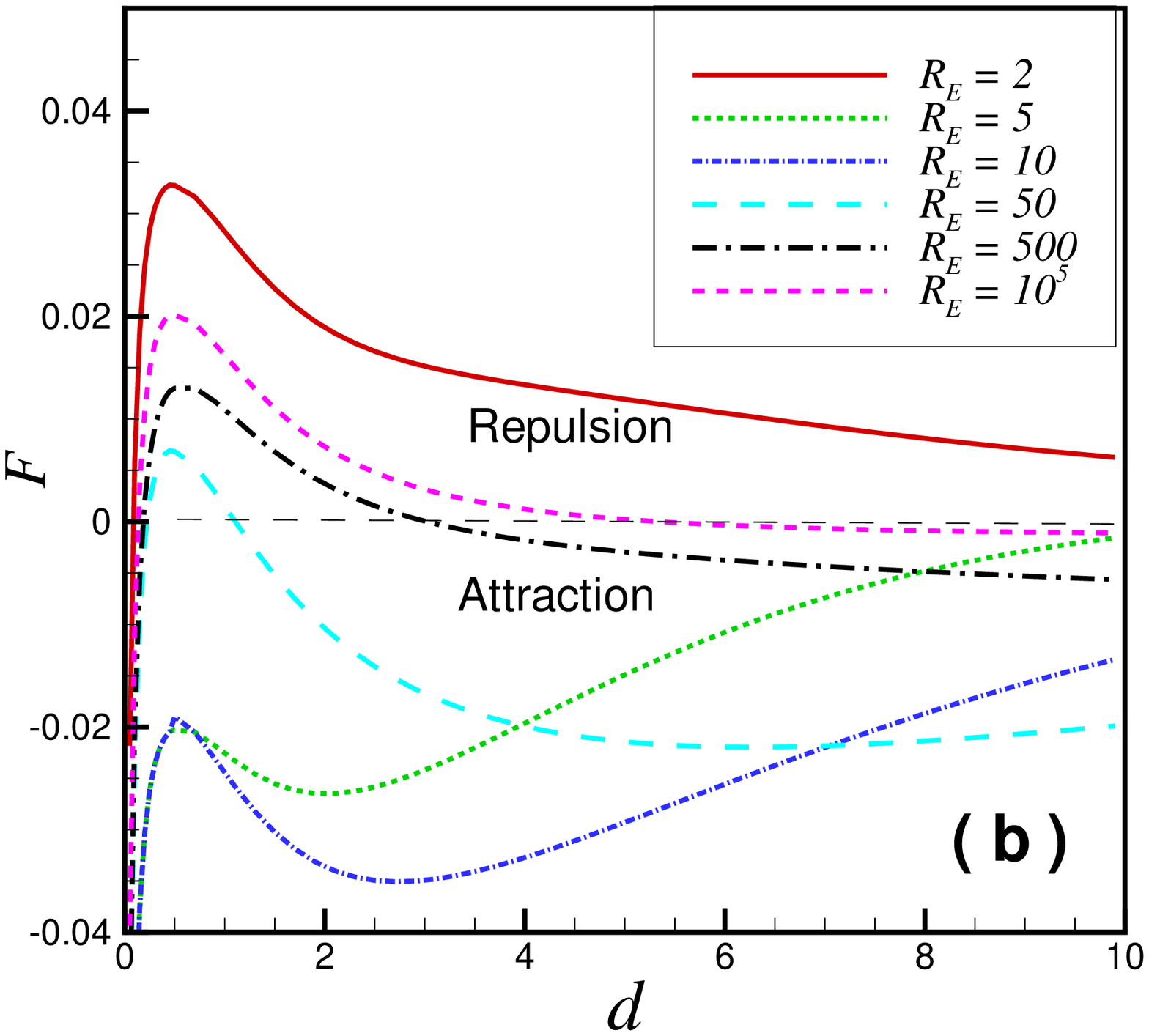} \\  \includegraphics[scale=.3]{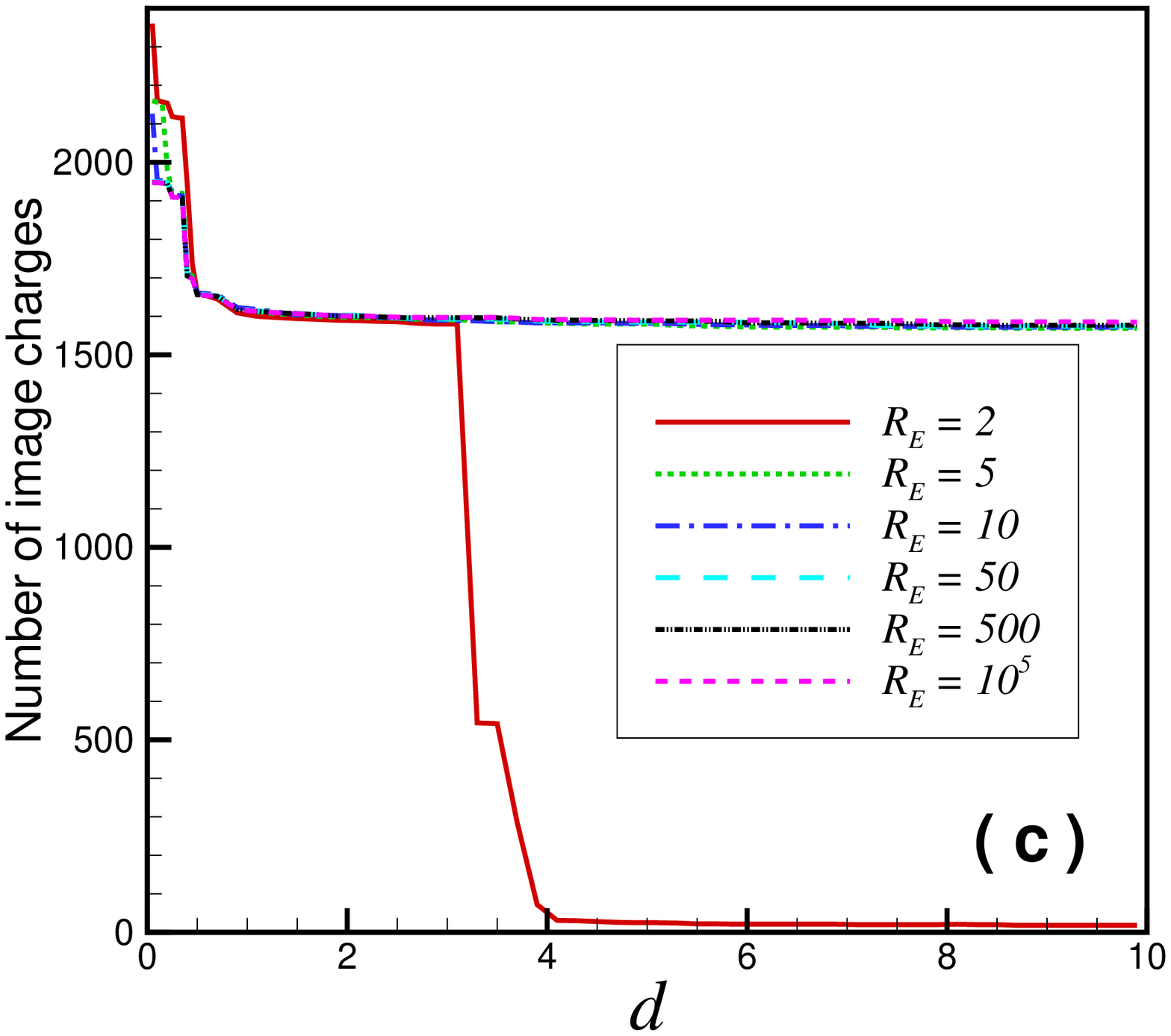}
\includegraphics[scale=.3]{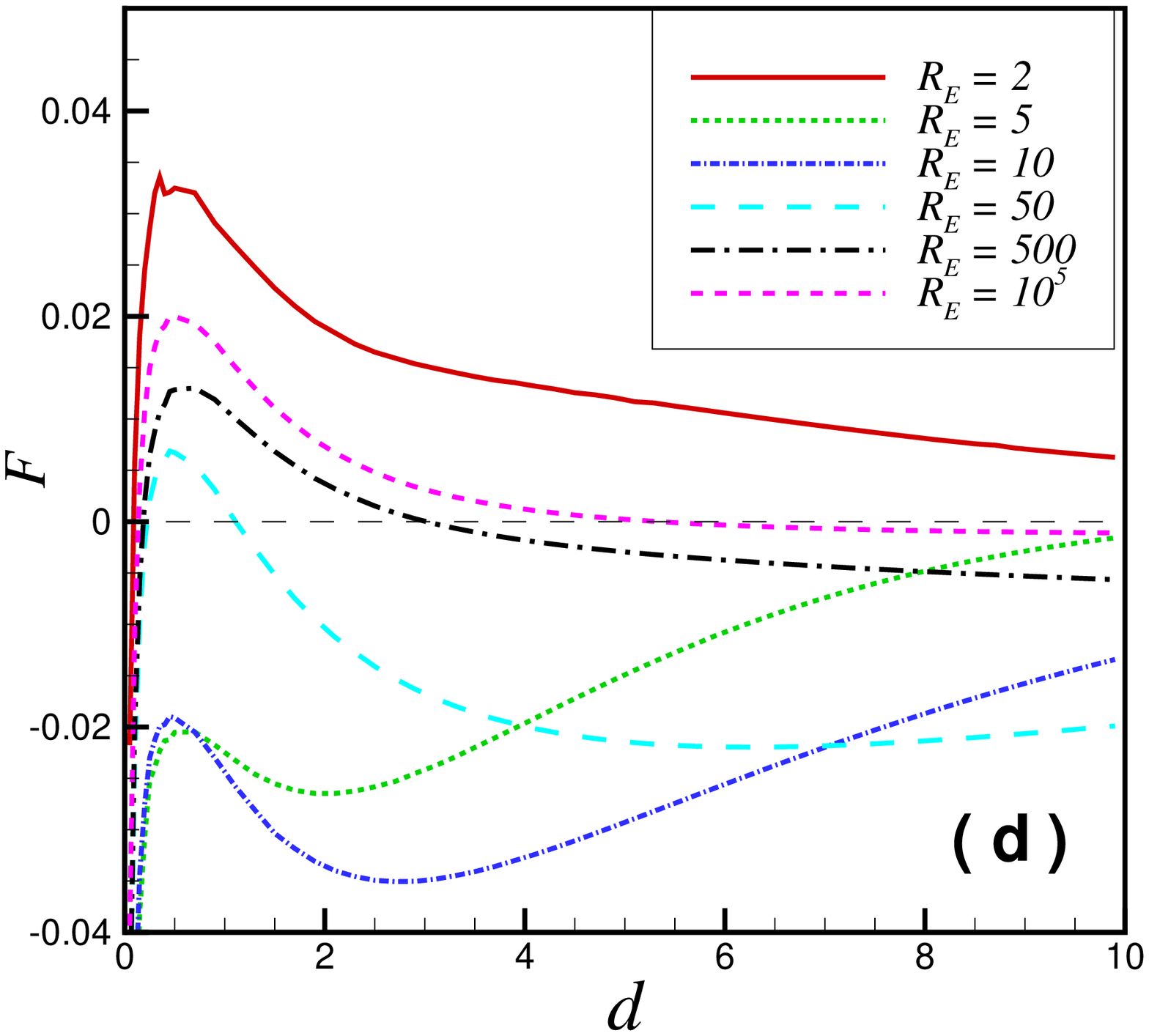}
\caption{Electrostatic interaction between like charged spheres near a neutral interface. $\varepsilon=80$ for all three objects (two charged spheres and a neutral large sphere).  (a) Schematic setting; (b) Interaction force between two charged spheres along the radius as a function of separation $d$ with $L5M3$ for image charges.  (c) Mean number of image charges with the coarse-graining algorithm of parameter $\delta=0.005$. (d) The interaction force with the coarse-graining treatment for image charges.} \label{threesph}
\end{figure*}

The polarization effect is essentially a many-body interaction. The presence of the third particle affects the interaction force between two charged spheres.
We study the interaction of two charged spheres with $R=1$ and $Q=1$ in the presence of an interface of different curvatures, i.e., another dielectric sphere with varying radius. The external sphere has a factor of $R_E$ of the size $R_1$, so there are three dimensionless distances, $d$, $R$ and $R_E$. The sphere is placed at at a distance $R+R_E$ away from the middle of the charged spheres; See Figure \ref{threesph} (a). The external sphere is charge-neutral, but with the same dielectric contrast to the background, $\varepsilon$, as the charged spheres. As the interactions with low dielectric ratios are always predicted repulsive, we focus on a dielectric permittivity $\varepsilon=80$, which models water droplets in the air \cite{DDL:JPCM:12}.

From Figure \ref{threesph} (b), it is observed the properties of these two spheres at a specified separation strongly and non-monotonously depend on the size of the external interface. Interestingly, all curves show attraction at small separation. When the separation is at the range of spherical radius, both large and small $R_E$ lead to repulsive forces ($R_E=2$ and $R_E\ge50$), and the intermediate radii ($R_E=5$ and 10)  are attractive. Starting from $d=0.5$, most of curves predict attractive forces for a wide range of $d$; but for small $R_E=2$ they are always repulsive. It could be interesting that, at $R_E=50$ and 500, the two particles have firstly an attraction and then an obvious repulsive force, and then they are attractive which continues for a long range. Finally, all cases are repulsive at the far field (not shown in the figure) as the polarization force is essentially a dipole interaction, shorter than the direct Coulomb repulsion. At the planar limit of the interface, $R_E=10^5$, the charged spheres are always repulsive (except the short separation since the surface charge distribution is not completely symmetric), in agreement with existing study \cite{Neu:PRL:99}.

We use this model to validate the coarse-graining algorithm again. In Figure \ref{threesph} (c) and (d), the results are the image number and interaction forces for the method with the coarse graining with the parameter $\delta=0.005,$ and we see Figure \ref{threesph} (d) accurately reproduce the results without using the coarse graining. For problems with three dielectric spheres, the total number of image charges with $L5M3$ for one source charge is 13995 by Eq. \eqref{imagenumber}. Shown in Figure \ref{threesph} (c), the numbers of images are generally around 1550 for $R_E\ge5$ and when the separation is not extremely small. The reduction is significant, but it is still high because most of image charges in the large sphere can not be coarse-grained. If the size of the external sphere is not so large, for example, in the case of $R_E=2$, the number becomes less than 30 when $d>4.5.$

\section{Conclusions}

In summary, we have investigated the image charge algorithm for evaluating electrostatic interaction between multiple dielectric spheres and reported the results for pairwise interactions of charged spheres in different dielectric environments. It is shown the algorithm is attractive to produce an accurate approximation of the polarization potential. And it is found that the like charge attraction is a general phenomenon for a wide range of dielectrics in condition of breaking symmetries such as size and charge asymmetry, and in the presence of an external interface.

The main objective of this paper is the design and optimization of the algorithm. Although it is superior to existing algorithms such as bi-spherical harmonics expansion \cite{RL:JCP:08} and boundary integral algorithms \cite{BGNHE:PRE:04,BGBEF:PRE:12} for multi-spheres, when the algorithm is applied to practical simulations, the computation of source-image interactions will be still expensive. By using the optimized image approximation, we are working on MC simulations of colloidal systems with two macroions to understand the role of image charges for colloidal suspensions at room temperature by taking account of both the ion correlation and the entropy contribution \cite{WBBP:JCP:99} and the effect of size and charge asymmetries \cite{MJL:CSA:11}. Because the polarization potential can be treated with point image charges, another important direction to speed up calculations is to develop order $N\log N$ algorithms for the pairwise interactions, and thus larger particle systems can be tackled.

\section*{Acknowledgements}

The author thanks Z. Gan, C. Holm and X. Xing for valuable discussions, and the referees for helpful suggestions. This work was supported by the Natural Science Foundation of China (Grant Numbers: 11101276 and 91130012). The author acknowledges the financial support from the Alexander von Humboldt foundation for a research stay at the Institute for Computational Physics, University of Stuttgart.


\end{document}